\documentclass[aps,twocolumn,amsmath,amssymb,showpacs,prl]{revtex4}
\usepackage{epsf}
\usepackage{graphicx}

\newcommand{\etal}{{\it et al.}}

\begin{document}

\title{Momentum anisotropy of the scattering rate in cuprate superconductors}
\author{
        A. Kaminski,$^{1,2,3}$
        H. M. Fretwell,$^{3}$
        M. R. Norman,$^{2}$
        M. Randeria,$^{4}$
        S. Rosenkranz,$^{1,2}$
        J. C. Campuzano,$^{1,2}$
        J. Mesot,$^{5}$
        T. Sato,$^6$ T. Takahashi,$^6$
        T. Terashima,$^7$ M. Takano,$^7$
        K. Kadowaki,$^8$
        Z. Z. Li$^9$
       and H. Raffy$^9$
      }
\affiliation{
         (1) Department of Physics, University of Illinois at Chicago,
             Chicago, IL 60607\\
         (2) Materials Sciences Division, Argonne National Laboratory,
             Argonne, IL 60439 \\
         (3) Department of Physics,University of Wales Swansea,
             Swansea SA2 8PP, UK\\
         (4) Tata Institute of Fundamental Research, Mumbai 400005,
             India\\
         (5) Laboratory for Neutron Scattering, ETH Zurich and PSI
             Villigen, CH-5232 Villigen PSI, Switzerland\\
         (6) Department of Physics, Tohoku University,
             980-8578 Sendai, Japan\\
         (7) Institute for Chemical Research, Kyoto University,
             Uji 611-0011, Japan\\
         (8) Institute of Materials Science, University of Tsukuba,
             Ibaraki 305-3573, Japan\\
         (9) Laboratorie de Physique des Solides,
                  Universite Paris-Sud, 91405 Orsay Cedex, France\\
         }
\date{\today}
\begin{abstract}
We examine the momentum and energy dependence of the scattering rate of 
the high temperature cuprate superconductors using angle resolved photoemission spectroscopy.
The scattering rate is of the form $a + b\omega$.  The inelastic coefficient $b$
is found to be isotropic.  The elastic term, $a$, however, is found to be
highly anisotropic in the pseudogap phase of optimal doped samples, with an
anisotropy which correlates with that of the pseudogap.  This can be contrasted
with heavily overdoped samples, which show an isotropic scattering rate in
the normal state.
\end{abstract}

\maketitle

There is a general consensus that understanding the normal state excitation
spectrum is a 
prerequisite to solving the high temperature superconductivity problem.
Angle resolved photoemission spectroscopy (ARPES) has played an important 
role in these studies because of the unique momentum and energy
resolved information it provides.  This includes the observation of dramatic
spectral lineshape changes caused by the 
superconducting transition \cite{NORM97},
the large momentum anisotropy of the superconducting gap 
consistent with d-wave symmetry \cite{SHEN93,DING96}, 
an anisotropic pseudogap above $T_c$ \cite{NAT96,LOESER}, and the
existence of nodal quasiparticles in the superconducting state \cite{ADAM00}
in contrast to marginal behavior seen above $T_c$ \cite{OLSON90,VALLA99}.
In this paper, we present data in the pseudogap and normal states of optimal and 
highly overdoped Bi2212 and Bi2201 compounds, and obtain the
energy and momentum dependence of the electron scattering rate.
We find that in optimal doped samples above $T_c$, the scattering rate 
is of the form $a + b\omega$, with the $b$ term isotropic.  The $a$ term, however,
is highly anisotropic, with a momentum dependence which follows that of the pseudogap.
In contrast, the highly overdoped samples have an isotropic scattering rate. 
The samples employed for this work are single crystals grown using the
floating zone method.  The optimal doped Bi$_2$Sr$_2$CaCu$_2$O$_{8+\delta}$
(Bi2212) samples ($T_c$=90K) were used in an earlier study \cite{ADAM01} as well
as the heavily overdoped ($T_c \sim 0$) Bi$_{1.80}$Pb$_{0.38}$Sr$_{2.01}$CuO$_{6-\delta}$
samples \cite{SATO01}. The optimally doped thin film samples of Bi$_{2}$Sr$_{1.6}$La$_{0.4}$CuO$_{y}$ were grown using an RF sputtering technique. The samples
were mounted with $\Gamma-M$
parallel to the photon polarization \cite{FOOT1} and cleaved in situ at pressures
less than 2$\cdot$10$^{-11}$ Torr.
Measurements were carried out at the
Synchrotron Radiation Center in Madison Wisconsin, on the U1 undulator
beamline supplying $10^{12}$ photons/sec,
using a Scienta SES 200 electron analyzer with a resolution in energy
of 16 meV and in momentum of 0.01 $\AA^{-1}$ for a photon
energy of 22 eV.

\begin{figure}
\includegraphics[width=3.4in]{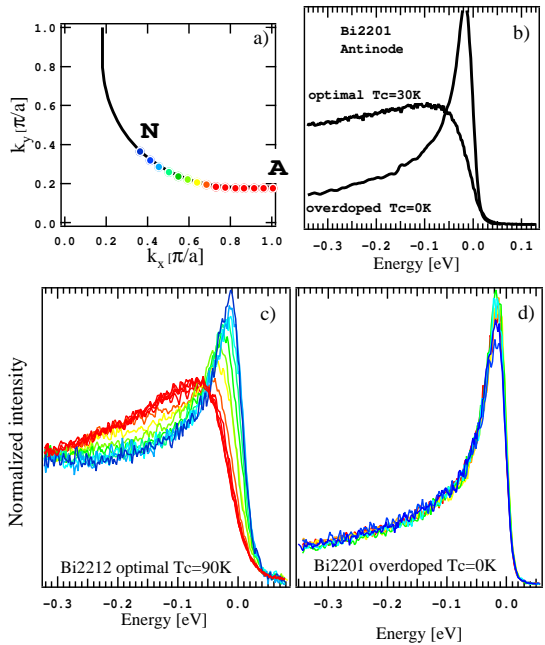}
\caption{Energy distribution curves (EDCs) along the Fermi surface: 
a) points on the Fermi surface where the EDC data were measured (N is the node,
A the antinode, of the d-wave gap).
%added temepratures
b) comparison of the data at the antinode for optimal doped and overdoped  Bi2201 obtained at T=50K and T=40K respectively.
c) EDC data from optimal doped Bi2212 ($T_c$=90K) at T=140K. 
The curves are color coded according to points in panel a). 
d) same data for overdoped Bi2201 ($T_c \sim 0$) at T=40K.  
}
\label{fig1}
\end{figure}

In Figs.~1c and 1d, we plot energy distribution curves (EDCs) along the 
Fermi surface (Fig.~1a) in the pseudogap state of optimal doped Bi2212 and the
normal state of highly overdoped Bi2201.
These data reveal that in the optimal doped case, there is a strongly
anisotropic pseudogap which is zero in an arc around the node of the
d-wave superconducting gap ($\Gamma-Y$ Fermi crossing), and takes its
maximal value at the antinode ($M-Y$ Fermi crossing).
Moreover, there appears to be a
strong anisotropy of the scattering rate, since the spectral peaks at 
the antinode are much broader than at the node.  Although this has been
suggested to be due to an unresolved energy splitting caused by bilayer
mixing \cite{BILAYER}, a recent study indicates that this is not the case for
optimal doped Bi2212 samples \cite{PRL03}.  Moreover, in Fig.~1b, we show data
at the antinode for optimal doped Bi2201, which has similar spectral
characteristics to that of Bi2212, again arguing against a bilayer effect.
We can contrast this behavior with that of heavily overdoped Bi2201
in the normal state, where no energy gap is present.  In this case,
the spectral peak is isotropic around the Fermi surface, indicating that 
the scattering rate is also isotropic. A similar conclusion was
reached in recent studies of heavily overdoped Bi2212 samples
where strong bilayer splitting is present \cite{BOG02}. 
\begin{figure}
\includegraphics[width=3.4in]{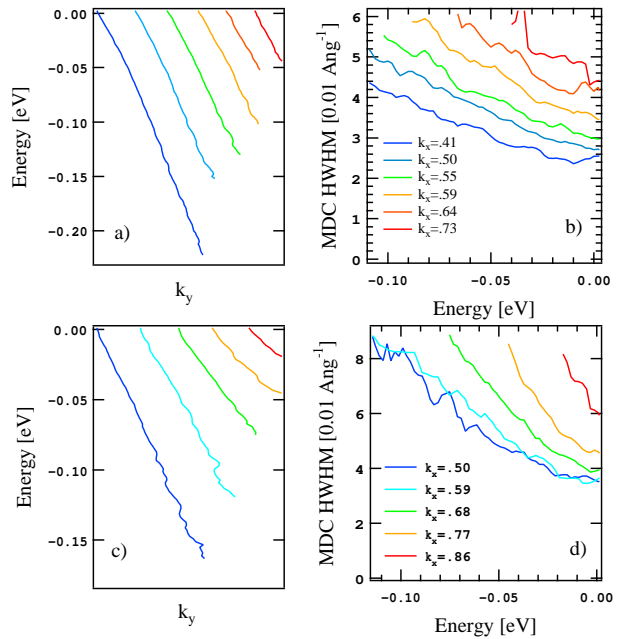}
\caption{Dispersion and peak widths obtained from momentum distribution
curves (MDCs) for selected momentum cuts parallel to $M-Y$ (labeled
by the $k_x$ value along $\Gamma-M$).
a), b) Bi2212 at T=140K and c), d) Bi2201 at T=40K.
}
\label{fig2}
\end{figure}

To obtain more quantitative information, we analyze momentum distribution
curves (MDCs) \cite{VALLA99}.  As we pointed out earlier \cite{ADAM01},
the MDC halfwidth (in the absence of an energy gap) is equal to the
imaginary part of the self-energy at that energy, $Im\Sigma(\omega)$, divided 
by the bare Fermi velocity, $v_{F0}$ (not the renormalized one, $v_F$).
In Fig.~2, we plot the
dispersion and peak widths obtained from MDCs along selected cuts in
momentum space parallel to the $M-Y$ direction.
The first interesting point to note is that the Fermi velocity (slope of the MDC
dispersion) appears to be fairly isotropic in optimal doped Bi2212, a
conclusion reached in an earlier ARPES study as well \cite{VALLA00}.
This is further quantified in Fig.~3d, where the velocity at the Fermi energy
is plotted around the Fermi surface.  This can be contrasted to heavily
overdoped Bi2201 (Figs.~2c and 3e), where the Fermi velocity is highly
anisotropic.  The latter result is consistent with a previous tight binding fit
to normal state ARPES dispersions in overdoped Bi2212 \cite{NORM95}.
Moreover, the velocity in optimal doped Bi2212 appears to increase slightly as the antinode
is approached, in contrast to the slight decrease observed by Valla \etal \cite{VALLA00}.
This difference occurs for two reasons.  First, the momentum cuts are parallel to $M-Y$.
Though these cuts are normal to the Fermi surface near the antinode,
a cut through the node is rotated 45 degrees from the normal.  As a consequence, the
Fermi velocity in the region near the node is underestimated by up to a factor of
$\sqrt{2}$.  Second, the appearance of a pseudogap in the spectrum will artificially inflate the value of the
Fermi velocity determined from MDCs for energies within the pseudogap \cite{MDC01}, since
the effect of the gap will cause the dispersion to become more vertical.  Thus,
the velocity is being overestimated in the region near the antinode.
Therefore, we conclude that the true Fermi velocity is roughly isotropic, with a
small decrease as the antinode is approached.  The strong
anisotropy observed in the Bi2201 case is due to the closeness
of the saddlepoint of the dispersion at $M$ (where the velocity is zero)
to the Fermi energy for this heavily overdoped sample \cite{SATO01}.
Stated another way, the small anisotropy in the optimal doped case implies
that the saddlepoint at $M$ is significantly far from the Fermi energy.

\begin{figure}
\includegraphics[width=3.5in]{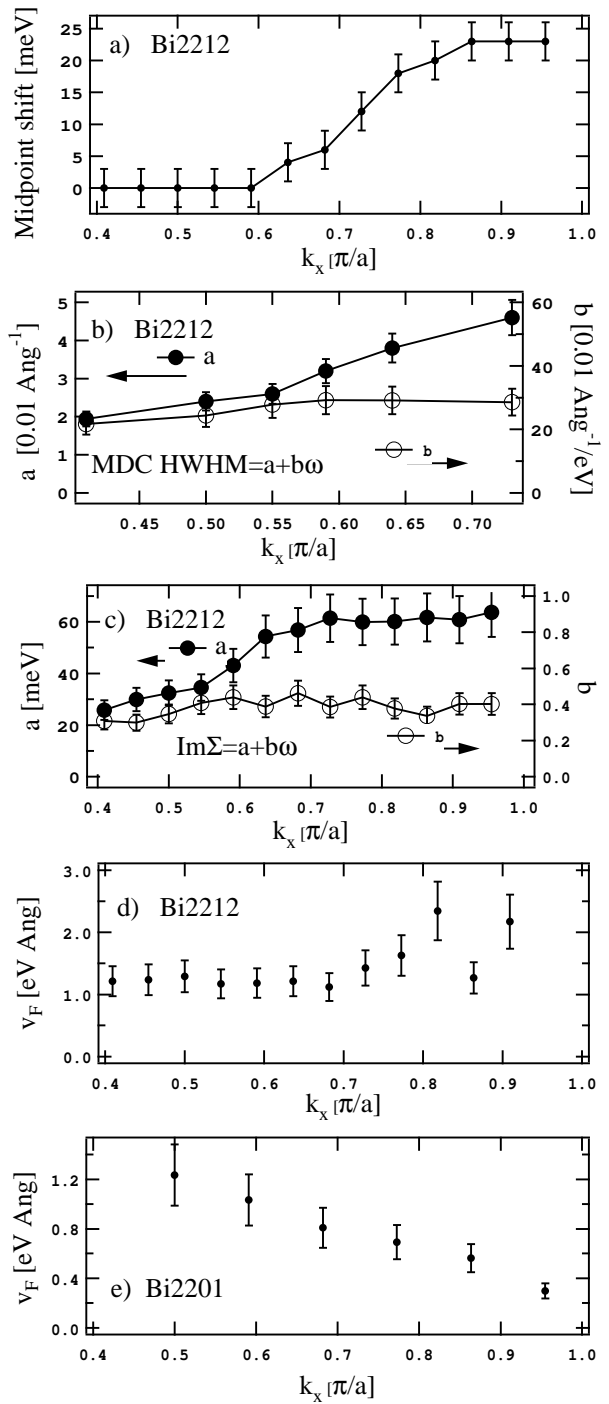}
\caption{a) Position of the midpoint of the leading edge of the EDC
around the Fermi surface
for Bi2212 obtained from Fig.~1c
($k_x$ labels the momentum cut as in Fig.~2, with $k_x=0.4$
corresponding to the node and $k_x=1.0$ to the antinode).  This is
an approximate measure of the pseudogap.
b) momentum dependence of the ``a'' (constant) and ``b'' (linear in $\omega$)
terms of MDC HWHM obtained by fitting data from Fig.~2b.
c) momentum dependence of the ``a'' (constant) and ``b'' (linear in $\omega$)
terms of $Im\Sigma$ obtained by fitting EDCs from Fig.~1c.
d) momentum dependence of the Fermi velocity (slope of the dispersion
in Fig.~2a).
e) the same as d), but for overdoped Bi2201 (Fig.~2c).}
\label{fig3}
\end{figure}

The energy dependence of the MDC peak widths for the various momentum cuts
is shown in Figs.~2b and 2d.  To a good approximation, the result for a particular
cut can be fit to the form $a + b\omega$.  This is analogous to the $a + bT$ form
indicated for the temperature dependence of MDC widths previously reported
by Valla \etal \cite{VALLA00}.   
In Fig.~3b, we show the momentum dependence of the $a$ and $b$ terms extracted 
from the MDC HWHM in Fig.~2b.  To complement these results, we have also fit EDCs 
along the Fermi surface using a model self-energy, $\Sigma$.  We have tested both 
quadratic and linear energy dependences for $Im\Sigma$ and found that only the latter 
is an adequate description of the data.  $Re\Sigma$ is determined by Kramers-Kronig 
transformation of $Im\Sigma=a+b\omega$, assuming the latter saturates at a constant 
value beyond a cutoff energy of 0.5 eV.  In Fig.~3c, we show the values of the $a$ and 
$b$ coefficients obtained from these EDC fits \cite{FOOT2}.

We first note that the $b$ term is isotropic in both plots.  Since the MDC HWHM is
equal to $Im\Sigma/v_{F0}$, this implies that the bare velocity ($v_{F0}$) is also 
isotropic.  The isotropy of $b$ provides strong support of the 
original marginal Fermi liquid conjecture \cite{MFL}.
At first sight, it would appear that the Bi2201 case is different, since the 
slope of the curves in Fig.~2d appears to increase as the antinode is
approached.  But once the velocity is divided out (Fig.~3d) \cite{FOOT3}, we find in this case as
well that the $b$ term for $Im\Sigma$ is isotropic, which is consistent with
the isotropy of the EDC lineshapes shown in Fig.~1.  We note that the $b$ term is
less well defined in the Bi2201 case due to curvature observed in Fig.~2d.  This is
expected, since as the hole doping increases, the lineshapes become more Fermi
liquid like, and thus one expects a crossover from linear to quadatic behavior in $\omega$.

The $a$ term in optimal doped Bi2212 (zero intercept in Fig. 2b) is found to be
highly anisotropic (Figs.~3b and 3c), as noted in the earlier study of Valla \etal \cite{VALLA00}.
This is consistent with the strong anisotropy of the EDC lineshapes shown in Fig.~1c.
Anisotropy in the ``zero intercept" is also evident in the heavily overdoped Bi2201 sample (Fig.~2d), but in this case, it can be accounted for by the anisotropy of the Fermi velocity, thus the $a$ term in $Im\Sigma$ in this case is isotropic (again, this
is consistent with the isotropy of the EDC lineshapes shown in Fig.~1d).  The anisotropy
of the $a$ term in the optimal doped sample has been attributed to off planar
impurities \cite{OFF}.  On the other hand, we note the remarkable similarity between
the anisotropy of this term (Fig.~3b) and that of the pseudogap (Fig.~3a).  This
indicates to us that the anisotropy is probably not due to impurity scattering, but rather a
consequence of the pseudogap.  This would be consistent with the observation of
isotropic lineshapes for more heavily overdoped samples of Bi2212 \cite{BOG02},
where no pseudogap is present.

One possibility for explaining this intriguing observation is to consider MDCs
in the presence of a spectral gap \cite{MDC01}.
At the Fermi energy ($\omega=0)$,
the resulting MDC HWHM is equal to $\sqrt{\Gamma^2+\Delta^2}/v_{F0}$,
where $\Gamma$ is the lifetime broadening and $\Delta$ the energy gap.
In the limit of small $\Gamma$, the anisotropy of the MDC HWHM follows that of $\Delta$ \cite{FOOT4}.
Thus, in this picture, we would interpret ``a" as containing a background contribution
due to instrumental resolution and weak impurity scattering, and an anisotropic term
due to the pseudogap.  But other possibilities could also be considered, such as the
cold spots model of Ioffe and Millis, where a highly anisotropic scattering rate is conjectured
due to scattering from d-wave pairing fluctuations \cite{IM,HUSSEY}.   

In conclusion, we find that the normal state scattering rate in the cuprates can be approximated by the form $a + b\omega$.  The inelastic $b$ term is found to be isotropic, which is a
necessary ingredient in the marginal Fermi liquid conjecture \cite{MFL}.  In contrast, the
$a$ term is found to be anisotropic for optimal doped samples, with the anisotropy
linked to that of the pseudogap.  For energies within the pseudogap, this term can be attributed to the influence of the gap on broadening the MDC linewidths, and thus should not be included in the ``normal" self-energy when analyzing transport data, a conclusion also reached by Millis and Drew \cite{MILLIS}.

This work was supported by the NSF 
DMR 9974401, the U.S. DOE, Office of Science, 
under Contract No. W-31-109-ENG-38 and the MEXT of Japan. The Synchrotron Radiation Center is 
supported by NSF DMR 9212658.  AK is supported by the Royal Society and Engineering and Physical Sciences Research Council, UK, 
and MR by the Indian DST through the Swarnajayanti 
scheme.  We acknowledge useful discussions with P.D. Johnson and C.M. Varma.


\begin{thebibliography}{99}

\bibitem{NORM97}
M. R. Norman \etal, Phys. Rev. Lett. {\bf 79}, 3506 (1997).

\bibitem{SHEN93}
Z. X. Shen \etal, Phys. Rev. Lett. {\bf 70}, 1553 (1993).

\bibitem{DING96}
H. Ding \etal, Phys. Rev. B {\bf 54}, R9678 (1996).

\bibitem{NAT96}
H. Ding \etal, Nature {\bf 382}, 51 (1996).

\bibitem{LOESER}
A. G. Loeser \etal, Science {\bf 273}, 325 (1996).

\bibitem{ADAM00}
A. Kaminski \etal, Phys. Rev. Lett. {\bf 84}, 1788 (2000).

\bibitem{OLSON90}
C. G. Olson \etal, PRB {\bf 42}, 381 (1990).

\bibitem{VALLA99}
T. Valla \etal, Science {\bf 285}, 2110 (1999).

\bibitem{ADAM01}
A. Kaminski \etal, Phys. Rev. Lett. {\bf 86}, 1070 (2001).

\bibitem{SATO01}
T. Sato \etal, Phys. Rev. B {\bf 64}, 054502 (2001).

\bibitem{FOOT1}
Zone notation is $\Gamma$ $(0,0)$, $M$ $(\pi,0)$, and $Y$ $(\pi,\pi)$ in units
of $1/a$, where $a$ is the Cu-Cu distance.

\bibitem{BILAYER}
D.L. Feng \etal, Phys. Rev. Lett. {\bf 86}, 5550 (2001);
Y.D. Chuang \etal, Phys. Rev. Lett. {\bf 87}, 117002 (2001).

\bibitem{PRL03}
A. Kaminski \etal, Phys. Rev. Lett. {\bf 90}, 207003 (2003).

\bibitem{BOG02}
P. V. Bogdanov \etal, Phys. Rev. Lett. {\bf 89}, 167002 (2002);
Z. M. Yusof \etal, Phys. Rev. Lett. {\bf 88}, 167006 (2002).

\bibitem{VALLA00}
T. Valla \etal, Phys. Rev. Lett. {\bf 85}, 828 (2000).

\bibitem{NORM95}
M. R. Norman, M. Randeria, H. Ding, and J. C. Campuzano, Phys. Rev. B {\bf
52}, 615 (1995).

\bibitem{MDC01}
M. R. Norman, M. Eschrig, A. Kaminski, and J. C. Campuzano, Phys. Rev. B
{\bf 64}, 184508 (2001).

\bibitem{FOOT2}
In these EDC fits, the pseudogap was taken into account in the simplest
possible manner by displacing the spectral peak to finite binding energy.

\bibitem{MFL}
C. M. Varma \etal, Phys. Rev. Lett. {\bf 63}, 1996 (1989).

\bibitem{FOOT3}
In the heavily overdoped case, where the renormalizations are weak and isotropic,
we can make the approximation of using the true Fermi velocity rather than the
bare one.

\bibitem{OFF}
E. Abrahams and C. M. Varma, Proc. Nat. Acad. Sci. {\bf 97}, 5714 (2000).

\bibitem{FOOT4}
This argument, though, would not explain the anisotropy of the EDC lineshapes at
high binding energies observed in Fig.~1c.

\bibitem{IM}
L. B. Ioffe and A. J. Millis, Phys. Rev. B {\bf 58}, 11631 (1998).

\bibitem{HUSSEY}
N. E. Hussey, Eur. Phys. J. B {\bf 31}. 495 (2003)

\bibitem{MILLIS}
A. J. Millis and H. D. Drew, Phys. Rev. B {\bf 67}, 214517 (2003).

\end{thebibliography}
\end{document}